# Two-section reactor model for autothermal reforming of methane to synthesis gas

P. Maarten Biesheuvel and Gert Jan Kramer

*Shell Global Solutions International B.V., Badhuisweg 3, 1031 CM, Amsterdam, The Netherlands.*

## Abstract

A one-dimensional and stationary reactor model is presented to describe the catalytic conversion of a gaseous hydrocarbon fuel with air and steam to synthesis gas by autothermal reforming (ATR) and catalytic partial oxidation (CPO). The model defines two subsequent sections in the reactor, namely an upstream oxidation section, and a downstream reforming section. In the oxidation section all of the oxygen is converted, with partial conversion of the fuel. An empirical fuel utilization ratio is used to quantify which part of the fuel is converted in the oxidation section as function of the relative flows of air and steam. In the oxidation section, the gas temperature rapidly increases toward the toptemperature at the intersection with the reforming section. In this section the temperature decreases while the fuel is further converted with water and $CO_2$ as oxidant. For methane as fuel, simulation results are presented and compared with experiments. For multicomponent fuels such as natural gas and naphtha, it is described how the two-section model can be applied.

## Introduction

Hydrogen is manufactured industrially from hydrocarbon fuels either by steam reforming or by gasification (Rostrup-Nielsen, 1993). In steam reforming, the fuel (in most cases, natural gas) reacts with steam toward synthesis gas over a Ni-based heterogeneous catalyst. Due to the endothermicity of the steam reforming reaction, heat must be supplied to the reactor through the reactor tube walls. Gasification is an autothermal process in which the fuel is mixed with pure oxygen and non-catalytically converted to synthesis gas. An example of its industrial use is the Shell Gasification Process (Elvers *et al.*, 1989).

The recent surge in demand for on-spot, small-scale, cheap and simple hydrogen production technologies based on hydrocarbon fuels is driven by the interest in fuel cells for electricity generation (Docter and Lamm, 1999; Ahmed and Krumpelt, 2001; Moon *et al.*, 2001). Light hydrocarbon fuels, such as natural gas, are envisioned for stationary applications, e.g., at the scale of a single household (1-5 kWe, kW electricity), while liquid fuels (gasoline, naphtha) have potential for automotive applications (25-75 kWe; Service, 1999). To obtain a gas mixture that can be fed to a fuel cell, the hydrocarbon fuel is first converted to synthesis gas in a reformer. When a proton exchange membrane (PEM) fuel cell is used, the synthesis gas must be further processed, first in a water-gas-shift reactor, where CO is converted to $CO_2$ and additional hydrogen is obtained, after which remaining traces of CO are removed (e.g., by selective oxidation) before the mixture is finally fed to the anode side of the fuel cell.

For these applications, down-scaled versions of the gasification and steam reforming processes are impractical. First of all, downscaling is not very well possible (especially for gasification) and otherwise the resulting processes are too bulky, expensive and difficult to operate. However, an air-based, autothermal catalytic fixed bed technology in which the hydrocarbon fuel is mixed with air and steam may result in a small reformer that is safe and easy to operate and automate. Two technologies are available, derived from their industrial analogues. Autothermal reforming (ATR), based on mixing air in with the steam/fuel mixture, is an autothermal adaptation of steam reforming. In the gasification process, it is possible to replace the oxygen with air and use a catalyst instead of an underoxidized burner; this is catalytic partial oxidation (CPO). Catalytic partial oxidation (CPO) has –within Shell– originally been developed as a catalytic alternative for the industrial SGP process. Originally intended for operation with pure oxygen, it has been modified for use with air and steam to suit the needs of fuel processing for fuel cells.

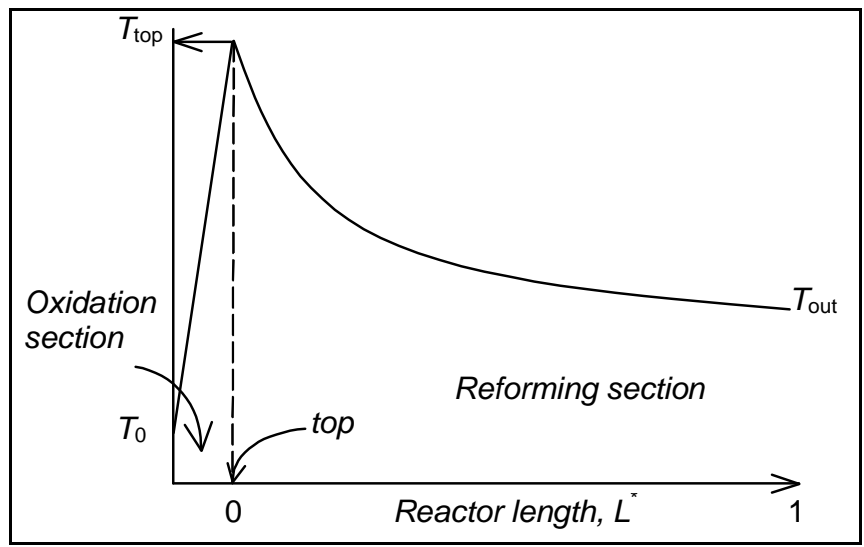

*Figure 1*. Gas phase temperature in CPO/ATR of methane. The measured *catalyst* temperature at the upstream side is assumed equal to the *gasphase* temperature at the intersection of the oxidation and reforming section (0, "top").

While the two technologies have a distinctly different origin, we are of the opinion that the two technologies have all but merged though CPO typically operates at low to zero steam-to-carbon ratio's (e.g., S/C<1) while ATR operates at a higher steam load (S/C>1). In ATR and CPO several similar catalyst systems have been used and tested, such as based on supported precious metals. In both processes the aim is to achieve the thermodynamic equilibrium composition which is determined by feed conditions (composition, temperature), pressure and heat loss. These conditions are optimized to obtain a maximum *syngas yield* in hydrogen and carbon monoxide (in the relevant yield figure, CO is added to $H_2$, because of its conversion in the subsequent water-gas-shift reactor), as well as to minimize slip of hydrocarbons (especially methane levels can be significant at thermodynamic equilibrium, dependent on exit temperature and amount of steam).

Comparing CPO with ATR, in CPO temperatures in the catalyst bed tend to be higher and syngas yield somewhat lower. However, premixing and preheating is more straightforward (certainly for S/C=0) and reaction rates tend to be higher (higher space velocity possible). In any case, the choice for the optimal air/fuel and steam/fuel ratio as well as inlet temperature and operating pressure must be made at the level of the entire fuel processor system.

It must be stressed that both in CPO and ATR the practical objective is to bring the mixture to thermodynamic equilibrium. While there is no a-priori reason why a catalytic process could not be more selective towards syngas than what is dictated by thermodynamics, the fact that CO and hydrogen are so much more reactive with oxygen than the feed hydrocarbon molecules makes this goal elusive. This reactivity bias will in fact cause an 'overshoot' of the reaction: excess amounts of $CO_2$ and water are produced at the "top" of the reactor (see *Figure 1*), after which thermodynamic equilibrium is approached via reforming reactions. Thus, if thermodynamic equilibrium is not reached, there is both an excess amount of $CO_2$ and water as well as an unnecessary slip of methane and higher hydrocarbons, both leading to a sub-optimal syngas yield.

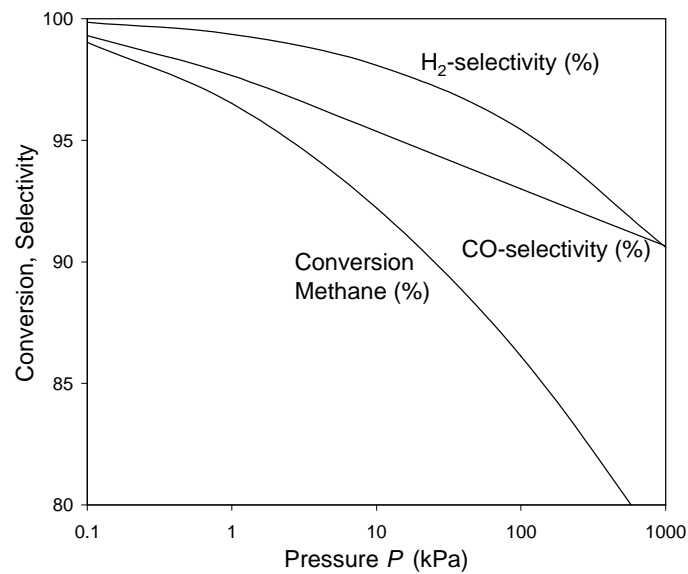

*Figure 2*. Thermodynamic equilibrium in adiabatic operation of air-based CPO of methane (no steam, $O_2/C$=0.5, air: 21 mol% $O_2$, 79% $N_2$), $T_{inlet}$=300 °C. Methane conversion and selectivities to CO and $H_2$ as function of pressure.

A second aspect of thermodynamic equilibrium is that for operation at low air and steam flows (relative to the fuel), such as done almost by definition in studies of "pure" CPO of methane (studies in which the $O_2/CH_4$ ratio is set to 0.5 and no steam is added, based on $CH_4$ + 0.5 $O_2$ → CO + 2 $H_2$), $CH_4$-conversion and CO- and $H_2$-selectivities are clearly below 100% unless operation is at a very high (inlet) temperature or a very low pressure (Dissanayake *et al.*, 1991; Wang and Ruckenstein, 1999). Results of thermodynamic calculations are shown in *Figure 2* for adiabatic CPO of methane without steam and with air ($O_2/CH_4$=0.5). In this case, only by decreasing the pressure to very low values, is it possible to obtain full methane conversion and 100% selectivity toward $H_2$ and CO. Thus, incomplete methane conversion and non-selective behavior at practical conditions (*T, P*) is not necessarily due to a catalyst with poor selectivity/activity, but in many cases a direct consequence of the thermodynamic equilibrium (Prettre, 1946; Chang and Heinemann, 1993; Bodke *et al.*, 1998). In practice this is well-known and to optimize syngas yield and keep methane slip below a certain threshold, more air (and/or water) are added than the "pure CPO" reaction suggests (e.g., Kramer *et al.*, 2001).

Thermodynamic equilibrium models required for the above calculations (both for CPO and ATR) are readily available or are rather easily set up, e.g., based on (part of) the equations used in this paper (Docter and Lamm, 1999). However, for reactor design and catalyst development, non-equilibrium reactor modeling is required.

This paper will focus on the development of a reactor model that describes the approach to thermodynamic equilibrium, thus reflecting the situation in practical reactors. Obviously, it is required that the model is thermodynamically consistent at each reactor coordinate (at each location). By this we mean that the equations that are used at each location to describe the conversion of the off-equilibrium gas mixture only contain local parameters (local temperature, gas composition) but correctly predict the thermodynamic equilibrium composition and temperature when given enough residence time or space. To be practical, the model must contain a limited number of fit parameters (reaction orders, activation energy, kinetic constants, etc.) Furthermore, the high upstream temperature (top-temperature) that is observed in CPO/ATR (Prettre *et al.*, 1946; Papp *et al.*, 1996; Heitnes Hofstad, 1997; Wang and Ruckenstein, 1999) must be an integral part of the model. Finally, atomic (mass) balances and the enthalpy balance must be explicitly accounted for.

Several relevant modeling efforts for CPO without steam (and some also for CPO/ATR with steam) can be found in literature (Hickman and Schmidt, 1993a,b; De Groote and Froment, 1995, 1996; Deutschmann and Schmidt, 1998; Arena et al., 2000; De Smet *et al.*, 2000, 2001), some based on effective rate equations, others on detailed descriptions of transport to the catalyst surface and reactions at the surface. Though several of these models contain some of the elements that are discussed above, we could not find a model that we could confidently use for our objectives.

Based on the above objectives, we set up a reactor model for synthesis gas formation in ATR and CPO processes. Our focus here is on methane, but it is explained how the model can be adapted to multicomponent fuels such as natural gas and naphtha. A one-dimensional and stationary (time-independent) model is set up that does not describe mass transfer to the catalyst surface or reaction at the catalyst surface in any detail but is based on effective rate equations expressed in gas phase properties (such as Eq. [10]) which makes it more straightforward to arrive at a thermodynamically consistent reactor model. For the existing models in literature that contain detailed descriptions of surface reactions in the form of reaction networks it is only possible to be thermodynamically consistent when enthalpy and entropy of each of the possible surface species are considered, when the enthalpy balance is expanded to include all of these surface species and when each adsorption, surface reaction and desorption step contains a driving force-term (such as the term within brackets in Eq. [9]) based on equilibrium of that specific step (the equilibrium constant in the driving force-term, $K_i$, is temperature-dependent, at least for each adsorption and desorption step).

In the model we only consider convective transport of mass and heat in the axial coordinate, which is possible because in CPO and ATR (both autothermal processes; in contrast to steam reforming) heat transfer in a perpendicular direction is negligible (only due to heat loss). Dispersive transport of mass and heat (in axial direction) are also neglected because of the high gas (space) velocities. We use a stationary model because it needs to be practical and we aim at describing stationary laboratory scale experiments and using the validated reactor model for design purposes. Instationary aspects are important during start-up, shut-down and load-change, but these periods are short in CPO and ATR that operate at a high space velocity.

The model divides the reactor into two subsequent sections. The first (upstream) section is the (partial) oxidation section where all oxygen is converted with part of the fuel and the highest "top" temperature is reached. This section is followed by the reforming section in which the remaining fuel is further converted with water and $CO_2$ as only possible oxidants to arrive at thermodynamic equilibrium, see *Figure 1* (Zhu *et al.*, 2001).

It is well-known that oxygen conversion (in the oxidation section) is very fast. Based on a kinetic analysis Zhu *et al.* (2001) estimate the oxidation section to be a few percent of the total reactor, while Kramer *et al.* (2001) estimate the oxidation section to comprise some ten percent of the total reactor volume, based on mass-transfer limited transport in the oxidation section. In the present model we do not assign reactor space to the oxidation section and focus on the reforming section.

At the end of the oxidation section (top), the crude syngas contains all components (CO, $CO_2$, $H_2$ and $H_2O$) as well as part of the fuel, but no oxygen. The fraction of the fuel that is converted in the oxidation section is the "fuel utilization ratio" (*UT*). A high fuel utilization ratio (for a given amount of air and steam) means that large amounts of CO and $H_2$ are formed in the oxidation section, resulting in low top-temperatures; a low value for *UT* results in high top-temperatures and large amounts of $CO_2$ and $H_2O$ formed in the oxidation section. In the latter case, more fuel is unconverted in the oxidation section and must be converted in the reforming section. For both reasons, a high value of *UT* is favourable for CPO/ATR performance. Note that the fuel utilization ratio is equal to one over the oxygen usage ratio in Kramer *et al.* (2001).

The fuel utilization ratio depends on the relative amounts of air and steam premixed with the fuel, and obviously depends on the rates by which the oxygen and the fuel are transported to the surface, adsorb and react. When the rate of transport and adsorption of the fuel (methane) is increased (relative to that of oxygen), the fuel utilization ratio increases (Kramer *et al.*, 2001).

The fuel utilization ratio is calculated from the catalyst temperature at the upstream side of the catalyst bed using an optical pyrometer and applying the appropriate mass and enthalpy balances (see Results and Discussion). When deriving the fuel utilization ratio from the measured temperature, it is assumed that at the end of the oxidation section the temperatures of gas and catalyst are equal (the "top-temperature") and that they coincide with the measured upstream catalyst temperature (see *Figure 1*). Our experimental observations indicate that this assumption is valid. Such a constant catalyst temperature throughout the oxidation section can be explained when the oxidation reaction is effectively heat- and mass-transfer limited, when (under CPO/ATR conditions) the Chilton-Colburn analogy holds and when the Lewis number (the dimensionless ratio between mass and heat diffusivity) is equal to unity. This case is argued further in Kramer *et al.* (2001). The assumption that the gas phase has the same temperature as the catalyst at the intersection of oxidation and reforming section (at the top, see *Figure 1*) is reasonable given the high heat and mass transfer rates, but can also be underpinned by the fact that in the oxidation section the gas is colder than the catalyst (the exothermic reactions take place on the catalyst surface while the gas is heated up by hot molecules coming off the surface), while in the (endothermic) reforming section the catalyst should be (somewhat) cooler than the gas. At the intersection of the exothermic and the endothermic regions the temperature of the catalyst and the gas phase should be the same.

In the model it is assumed that the water-gas-shift (WGS) equilibrium is established at each location in the reforming section (thus also at the top where the reforming section starts). That the WGS-equilibrium is established is quite often reported in work on steam reforming and dry reforming (Bodrov *et al.*, 1964; Soliman and Adris, 1992; Papp *et al.*, 1996; Hou and Hughes, 1997). Also, in experiments with partial fuel conversion (but complete oxygen consumption) we find that the syngas from the reformer is at or close to the WGS-equilibrium. It must be noted that assuming that the WGS-equilibrium is established is not a critical assumption in the current model because of the following reasons: the reaction enthalpy of the WGS-reaction is not very large (small influence on enthalpy balance) and the reaction rate equation that is used for the fuel in the reforming section, as well as the syngas yield-number, are independent of whether the WGS-reaction is at equilibrium or not (for the syngas yield, CO and $H_2$ are added together).

In the Theory section the model for CPO and ATR is explained in detail for methane as fuel, and more schematically for natural gas and naphtha. In the Results and Discussion section, simulation results are presented that are representative of our experience for CPO and ATR of methane.

## Theory

Inputs

The two-section model describes the catalytic conversion of methane mixed with air and steam. The inlet molar flows, $\phi_{CH4,0}$, $\phi_{air,0}$, $\phi_{water,0}$ [mol/s] are known as well as the inlet temperature, $T_0$ [K] and the system pressure, $P$ [Pa] (absolute pressures are used throughout). It is assumed here that the fraction of $O_2$ in air is $\chi$ (namely, $\chi=0.2095$), the remainder being $N_2$, thus $\phi_{O2,0}=\chi\cdot\phi_{air,0}$ and $\phi_{N2,0}=(1-\chi)\cdot\phi_{air,0}$. The humidity of air is neglected as well as Ar, $CO_2$, etc. The oxygen-to-carbon ratio is $\alpha=\phi_{O2,0}/\phi_{CH4,0}$ and the steam-to-carbon ratio is $\gamma=\phi_{H2O,0}/\phi_{CH4,0}$. Simulations are based on a fuel flow of $\phi_{CH4,0}=10$ mmol/s which corresponds to 8 kWh (heating value) and ~3 kWe produced in the fuel cell, which is typical of the scale of a single household.

The reactor model consists of a description of the oxidation section in which all oxygen is converted as well as part of the methane. This conversion is considered to be extremely fast, thus the catalyst volume required negligible. The exit of the oxidation section is called the "top" having a certain gas composition and a top-temperature. The top is the virtual intersection of the oxidation section and the reforming section. In the model, the entire catalyst volume is assigned to the reforming section.

Heat loss can be implemented in both sections. For the oxidation section it must be an algebraic expression (as function of top-temperature, surface area, etc.) or a constant number, $Q_{ox}$ [W]. Heat loss in the reforming section is a function of the local temperature and is implemented in the (differential) enthalpy balance, Eq. [12].

Simple relations

Some simple relations that will be used without further reference are summarized here. Molar flows $\phi_i$ [mol/s] are related to the total flow, $\phi_{tot}$ by $\phi_i = x_i \cdot \phi_{tot}$, with $x_i$ the molar fraction [mol/mol]. Thus, $\phi_{tot} = \sum_i \phi_i$ and $\sum_i x_i = 1$. At the inlet, $i$ runs over max. 4 species (CH$_4$, O$_2$, N$_2$ and possibly H$_2$O) and over 6 species at the top and in the reforming section (CH$_4$, CO, CO$_2$, H$_2$, H$_2$O and N$_2$). Partial pressures $p_i$ [Pa] are related to the total pressure $P$ by $\sum_i p_i = P$ and $p_i = x_i \cdot P$.

Thermodynamics

The relevant thermodynamics that is included in the oxidation section and reforming section is summarized in a general form. Enthalpy, $H_i$ [J/mol], and entropy, $S_i$ [J/(mol·K)], of a certain component, $i$, are a function of the local temperature, $T$, see *Appendix A*.

The two relevant reactions are the water-gas-shift (WGS) reaction,

CO + H$_2$O $\longleftrightarrow$ CO$_2$ + H$_2$ ,

and the methane-steam-reforming (MSR) reaction,

CH$_4$ + H$_2$O $\longleftrightarrow$ CO + 3 H$_2$ .

For each reaction, the equilibrium constant, $K_i$, is given by

$$K_i = \exp\left(-\frac{1}{RT}\left(\sum_i \nu_i H_i - T\sum_i \nu_i S_i\right)\right) = \prod_i \left(\frac{p_i}{P^*}\right)^{\nu_i} \quad [1]$$

with $\nu_i$ the stoichiometric constants in the reaction and $P^*$=101.3 kPa. For the two reactions the result is

$$K_{WGS} = \exp\left(-\frac{1}{RT}\left((H_{H2} + H_{CO2} - H_{CO} - H_{H2O}) - T(S_{H2} + S_{CO2} - S_{CO} - S_{H2O})\right)\right) =$$
$$= \frac{p_{H2} \cdot p_{CO2}}{p_{H2O} \cdot p_{CO}} = \frac{x_{H2} \cdot x_{CO2}}{x_{H2O} \cdot x_{CO}} = \frac{\phi_{H2} \cdot \phi_{CO2}}{\phi_{H2O} \cdot \phi_{CO}} \quad [2]$$

$$K_{MSR} = \exp\left(-\frac{1}{RT}\left((3 \cdot H_{H2} + H_{CO} - H_{CH4} - H_{H2O}) - T(3 \cdot S_{H2} + S_{CO} - S_{CH4} - S_{H2O})\right)\right) =$$
$$= \frac{p_{H2}^3 \cdot p_{CO}}{p_{H2O} \cdot p_{CH4}}\left(\frac{1}{P^*}\right)^2 = \frac{x_{H2}^3 \cdot x_{CO}}{x_{H2O} \cdot x_{CH4}}\left(\frac{P}{P^*}\right)^2 = \frac{\phi_{H2}^3 \cdot \phi_{CO}}{\phi_{H2O} \cdot \phi_{CH4}} \frac{1}{\phi_{tot}^2}\left(\frac{P}{P^*}\right)^2 \quad [3]$$

Atomic balances are based on the fact that at stationary conditions, and for a one-dimensional reactor model, at each location, the mole flow of each atom (C, H, O, N) is constant (thus equal to the inlet, 0, conditions). Thus, at each location, $j$, the balances for C, H, O and N are

$$\phi_{CH4,0} = \phi_{CH4,j} + \phi_{CO,j} + \phi_{CO2,j}$$
$$\phi_{CH4,0} \cdot (4 + 2 \cdot \gamma) = 4 \cdot \phi_{CH4,j} + 2 \cdot (\phi_{H2,j} + \phi_{H2O,j})$$
$$\phi_{CH4,0} \cdot (2 \cdot \alpha + \gamma) = \phi_{CO,j} + 2 \cdot \phi_{CO2,j} + \phi_{H2O,j}$$
$$\phi_{N2,0} = \phi_{N2,j}$$
[4]

Oxidation Section

In the oxidation section, kinetics are not considered but a set of algebraic equations describes the conditions at the exit of the oxidation section (the top) which is the intersection with the downstream reforming section. The top-temperature, $T_{top}$, and the individual gas flows ($\phi_{CH4,top}$, $\phi_{CO,top}$, $\phi_{CO2,top}$, $\phi_{H2,top}$, $\phi_{H2O,top}$, $\phi_{N2,top}$) are unknown. At the end of the oxidation section oxygen has already vanished. Seven equations are required. We have the WGS-equilibrium, Eq. [2], the four (atomic) mass balances, Eq. [4], and the enthalpy balance

$$\sum_i H_{i,T_0} \cdot \phi_{i,0} = \sum_i H_{i,T_{top}} \cdot \phi_{i,top} + Q_{ox}$$
[5]

with $Q_{ox}$ the heat loss in the oxidation section [W], which can be made an explicit function of $T_0$, $T_{top}$, surface area, etc.

Finally, we introduce the fuel-utilization ratio, $UT$, which describes which part of the fuel is converted in the oxidation section. $UT$ is a function of process conditions, catalyst formulation and structure. For a specific catalyst we have derived the empirical function

$UT = -1.6492 \cdot \alpha^2 + 2.664 \cdot \alpha - 0.2348 - 0.033 \cdot \gamma$ [6]

from experiments with methane at $T_0 \sim 400\ ^oC$ and $P=170$ kPa in a wide range of values for $\alpha$ and $\gamma$ (being respectively the oxygen-to-carbon and steam-to-carbon ratio). In these experiments the top-temperature was measured and the above model used to derive the appropriate $UT$-value. Eq. [6] is a very accurate description of these data.

Using Eq. [6], $\phi_{CH4,top}$ is given by

$$\phi_{CH4,top} = (1 - UT) \cdot \phi_{CH4,0}.$$
[7]

Note that this set of equations does not explicitly describe the selectivities to $H_2$ and CO (relative to the formation of $H_2O$ and $CO_2$). However, selectivities can be directly calculated from the oxidation-model (based on Eq. [6], atom and enthalpy balances and the WGS-equilibrium). In general, (for a given $\alpha$ and $\gamma$), the higher $UT$, the more selective the oxidation section and the lower the top-temperature.

Reforming section

Transport and reaction in the reforming section are described by a plug-flow model. The four atomic balances are used as well as the enthalpy balance and the WGS-equilibrium. One differential mass balance is required to obtain a one-dimensional, stationary, plug-flow model for the reforming section. Ideally, this is a mass balance in one of the rate-limiting reactants, typically in the fuel component (but not necessarily so). Indeed, we will use a differential mass balance that describes the conversion of methane. Based on literature (Akers and Camp, 1955; Bodrov *et al.*, 1964; Ridler and Twigg, 1989; Rostrup-Nielsen, 1993; Steghuis *et al.*, 1998) and our own experience, a power law rate equation is proposed for the reforming of methane that is first order in the partial pressure of the fuel component, methane. A first order rate equation in the fuel is compatible with the idea that mass transfer, adsorption or dissociation of the fuel is rate limiting.

At stationary conditions, a differential mass balance over a "slice" in the reforming section results for component $i$ in

$$\frac{d\phi_i}{dz} = R_i \cdot S. \qquad [8]$$

with $S$ the cross-sectional surface area [m$^2$], $z$ the reactor coordinate [m], $\phi_i$ the mole flow of $i$ [mol/s] and $R_i$ the formation rate of $i$ [mol/(m$^3\cdot$s)].

For methane conversion, the following effective rate equation is proposed based on a first order dependence on the partial pressure of methane and an empirical term (in between brackets) that is based on local partial pressures and (via $K_{MSR}$) on the local temperature, which describes how closely thermodynamic equilibrium has been approached. The result is

$$R_{CH4} = -a \cdot k_0 \cdot e^{-E_A/RT} p_{CH_4} \left(1 - \frac{p_{H2}^3 \cdot p_{CO}}{K_{MSR} \cdot p_{CH4} \cdot p_{H2O}} \frac{1}{P^{*2}}\right) \qquad [9]$$

in which $a$ is the specific surface area $a$ (m$^2$ of active phase per m$^3$ of reactor volume) which is a function of the amount of active phase and details of catalyst preparation, form and structure; $k_0$ is the pre-exponential kinetic number, $E_A$ the activation energy [J/mol] of the reaction and $R$ the gas constant [J/(mol·K)]. The activation energy of reforming reactions is often reported to be in the 100 kJ/mol-range: Hou and Hughes (1997) suggest $E_A$=100 kJ/mol (based on their Fig. 4) and $E_A$=95 kJ/mol (Fig. 12), while Steghuis et al. (1998) measure $E_A$=105-114 kJ/mol. $K_{MSR}$ is a direct function of the local temperature and is given by the first equality in Eq. [3].

Because the WGS-equilibrium is assumed to be established throughout the reforming section, the driving force-term (between brackets) based on the MSR-reaction, can be rewritten to a form based on the dry reforming (DR) equilibrium (CH$_4$+CO$_2$ <--> 2·CO+2·H$_2$) or to a form based on any of the other possible reactions of methane with water and CO$_2$. All give exactly the same result.

Eq. [9] is implemented in Eq. [8] while partial pressures are recalculated to flows, $\phi_i$. Furthermore, $z$ and $S$ are replaced by the catalyst volume, $V$ [m$^3$] and a dimensionless reactor coordinate, $0<L^*<1$, $z\cdot S=L^*\cdot V$. Finally, the group $a\cdot k_0 \cdot V$ is replaced by $\lambda$ [mol/(Pa·s)], resulting in

$$\frac{d\phi_{CH4}}{dL^*} = -\lambda \ e^{-E_A/RT} p_{CH4}\left(1 - \frac{p_{H2}^3 \cdot p_{CO}}{K_{MSR} \cdot p_{CH4} \cdot p_{H2O}} \frac{1}{P^{*2}}\right) \ \rightarrow$$

$$\frac{d\phi_{CH4}}{dL^*} = -\lambda \ e^{-E_A/RT} P\phi_{CH4}\phi_{tot}^{-1}\left(1 - \frac{\phi_{H2}^3 \cdot \phi_{CO}}{K_{MSR} \cdot \phi_{CH4} \cdot \phi_{H2O}}\left(\frac{P}{P^*}\right)^2 \frac{1}{\phi_{tot}^2}\right). \qquad [10]$$

Note that Eq. [11] does not directly specify whether methane reacts with water or CO$_2$, nor does it specify the products; it only describes the conversion rate of methane. However, in conjunction with the WGS-equilibrium and the atom balances, Eqs. [4], the conversion rates of all other components are fixed at each location, $L^*$.

Finally, the differential enthalpy balance is given by

$$\frac{d\sum_i H_i \phi_i}{dz} = -U \cdot P_R \cdot (T - T^*).\qquad [11]$$

Here, an engineering heat transfer relation is used for heat loss through the wall of the reactor. $U$ is an effective heat transfer coefficient [W/(m$^2$·K)], $T^*$ a representative temperature outside the reactor, $P_R$ the perimeter [m]. Multiplying each side by $V/S$ and replacing $U \cdot P_R \cdot V/S$ by $\kappa$ [W/K] results in

$$\frac{d\sum_i H_i \phi_i}{dL^*} = -\kappa \cdot (T - T^*).\qquad [12]$$

At each location, that is at each $L^*$-value, the two differential equations, Eqs. [10] and [12], are solved as well as the four atom balances, Eqs. [4], the WGS-equilibrium, Eq. [2], and the first equality in Eq. [3]. Because the temperature, $T$, changes as function of $L^*$, all terms $H_i$, $S_i$, $K_{WGS}$ and $K_{MSR}$ change with $L^*$, as well as all $\phi_i$'s and $\phi_{tot}$; $P$, $\kappa$, $\lambda$ and $T^*$ are constant throughout.

Coke formation

From separate thermodynamic calculations one can derive the following inequality that predicts below what critical temperature, $T_{critical}$ [K], coke formation is thermodynamically expected,

$$T_{critical}=3.1368 \cdot (\ln(A))^2 - 45.791 \cdot \ln(A) + 973.51 \text{ with } A = \frac{p_{CO2}}{p_{CO}^2} P^* = \frac{x_{CO2}}{x_{CO}^2}\frac{P^*}{P} = \frac{\phi_{CO2}}{\phi_{CO}^2}\phi_{tot}\frac{P^*}{P}.\qquad [13]$$

After having determined the thermodynamic equilibrium gas phase composition assuming zero coke formation, Eq. [13] can be used to check whether that assumption was correct. Coke formation is not predicted in the relevant cases for ATR/CPO we considered, with the equilibrium temperature of the synthesis gas typically ~100 K above that required for coke formation. This is in alignment with our experiments in which coke formation in the reactor is not observed under normal operating conditions.

Natural gas

The model for ATR/CPO of methane can be extended to describe the conversion of natural gas - if methane makes up the majority of the hydrocarbons in the natural gas, which is generally the case. The two-section model for methane can be used as long as we assume that all hydrocarbons except for CH$_4$ are consumed in the oxidation step (Akers and Camp, 1955). The parameters $\alpha$ and $\gamma$ required in Eq. [6] are best defined relative to the total inlet hydrocarbon flow, $\phi_{C,0}$: $\alpha=\phi_{O2,0}/\phi_{C,0}$ and $\gamma=\phi_{H2O,0}/\phi_{C,0}$ with $\phi_{C,0}$ given by a summation over all hydrocarbons

$$\phi_{C,0} = \sum_i C_i \cdot \phi_{C_i,0} \qquad [14]$$

with $C_i$ the $C$-number of component $i$ (e.g., $C=2$ for C$_2$H$_6$). Further, Eq. [7] must be replaced by

$$\phi_{CH4,top} = (1-UT) \cdot \phi_{C,0}.\qquad [15]$$

<u>Naphtha and other multicomponent fuels</u>

Naphtha is a complex mixture of components mainly in the $C_5$-$C_{12}$ range. Instead of dealing with each component separately, it is possible to simplify the analysis by proper averaging. For example, a fuel mixture of 25 mol% toluene, 25% methylcyclohexane, 25% n-heptane and 25% iso-octane can be summarized by the formula $C_{7.25}H_{14}$ with the enthalpy function $H = -6 \cdot 10^{-5} \cdot T^3 + 0.2638 \cdot T^2 + 15.249 \cdot T - 153627$ (J/mol) correct within ~1% in the relevant temperature range 298<$T$ [K]<1400.

A reformer model for naphtha can be derived from the methane two-section model. For the oxidation section, a utilization-function for naphtha must be used which is not necessarily the same as Eq. [6], as well as an additional equation for the formation of methane and other hydrocarbons (though methane is the most relevant one; the formation rates of higher hydrocarbons are negligible with ppm- or ppb-levels in the equilibrium mixture). Methane formation in the oxidation section can be set to zero, $\phi_{CH4,top}=0$, or an equilibrium can be used, such as the MSR-equilibrium or the dry reforming equilibrium (DR).

In the reforming section a differential mass balance for naphtha conversion (e.g., first order in naphtha) replaces Eq. [11]. A driving force-term, such as used in Eq. [10] is not required, as at thermodynamic equilibrium naphtha-conversion is complete. For methane formation, the MSR- or DR-equilibrium can be used or a reaction rate equation, e.g., based on the methanation mechanism (with the rate dependent on the partial pressures of CO and $H_2$). We find that the DR-equilibrium is in many cases established (at the exit of the reactor) and use the DR-equilibrium in the reforming section.

## Results and Discussion

In the first part of Results and Discussion, experimental results of the conversion of methane in ATR/CPO are compared with predictions of the two-section model. In the second part somewhat other parameter settings are used and simulation results are presented for top-temperature and methane conversion behavior.

Two remarks must be made with respect to methane conversion, $\zeta$, and utilization, $UT$. First, whereas $\zeta$ is the "actual" conversion measured at the end of the reactor, thus at the end of the reforming section, the fuel utilization ratio, $UT$, is actually equal to the methane conversion at the end of the oxidation section (0<$UT$<1). The difference between $UT$ and $\zeta$ is a measure of the fuel conversion in the reforming section.

Second, in CPO/ATR of methane, the methane conversion, $\zeta$, is directly related to the syngas yield, $Y$, (mole flow of produced CO+$H_2$ divided by the mole flow of $CH_4$ fed to the reactor). This relation follows from combination of the atom balances for C, H and O, and is given by

$Y = 4 \cdot \zeta - 2 \cdot \alpha$ [16]

in which $\zeta$ is the methane conversion (0<$\zeta$<1) and $\alpha$ the (molecular) oxygen-to-carbon ratio. Thus, methane conversion results can be recalculated directly to syngas yield numbers. Eq. [16] can be used to calculate the actual syngas yield (based on $\zeta$), the syngas yield after the oxidation section (based on $UT$) and the maximum attainable syngas yield (based on the conversion at thermodynamic equilibrium).

Comparison of two-section model with experimental results

Experimental data for top-temperature and methane conversion are presented for a system pressure of 170 kPa and an inlet temperature of $T_0 \sim 400$ °C in *Figure 3* and *Figure 4* for different values of the oxygen-to-carbon ratio, $\alpha$, and the steam-to-carbon ratio, $\gamma$. Each experiment is analyzed individually and a different heat loss is calculated for each experiment by the difference of the total enthalpy leaving and entering the reactor (inlet temperature measured using a thermocouple; exit temperature by assuming the WGS-equilibrium; exit composition based on GC-analysis). In the analysis the heat loss is completely assigned to the oxidation section; therefore $Q_{ox}$ of Eq. [5] can be measured. In the analysis of the data (*Figures 3 and 4*) heat loss in the reforming section is set to zero ($U$ in Eq. [12] set to zero). Due to the fact that the heat loss is different for each experiment, the lines for toptemperature (*Figure 3*), conversion and thermodynamic equilibrium (*Figure 4*) are somewhat jagged. The heat loss was in all cases small compared to the total energy content (in combustion) of the entire gas flow.

From each experimental value of top-temperature (symbols in *Figure 3*) the utilization of methane ($UT$) can be calculated (circles in *Figure 4*) using the oxidation model that is based on atom balances, the enthalpy balance and the WGS-equilibrium (i.e., given these constraints, only for one value of $UT$ is the measured top-temperature, at the exit of the oxidation section, obtained). The top-temperature decreases when the amount of air is reduced or the amount of steam is increased. The empirical utilization function, Eq. [6], is derived from the derived utilization data and describes these data quite well, see the lower solid lines in *Figure 4*. Using Eq. [6] and the other elements of the oxidation model the top-temperature can be back-calculated. The result is plotted as the solid lines in *Figure 3*. For the highest oxygen-to-carbon ratio, $\alpha=0.59$, the prediction is accurate; for the two lower oxygen-to-carbon ratio's the prediction is less accurate but is certainly showing the right trends.

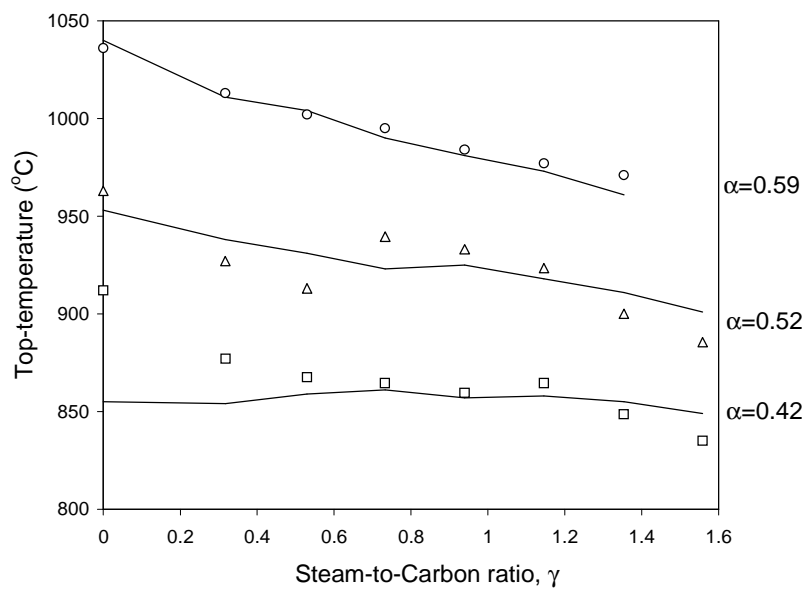

*Figure 3*. Top-temperature in ATR/CPO of methane. Different steam-to-carbon ratio's, $\gamma$, and oxygen-to-carbon ratio's, $\alpha$. $T_0 \sim 400$ °C, $P=150$ kPa, varying heat loss. Lines are predictions of the oxidation model.

The data for methane conversion, $\zeta$ (squares) are presented in *Figure 4* together with the predicted conversion at thermodynamic equilibrium (upper solid lines). By adjusting $\lambda$ and $E_A$ (given a suitable form of the rate equation, Eq. [10]), the two-section reactor model (dashed lines) can be fitted to the methane conversion data.

For each oxygen-to-carbon ratio, $\alpha$, the difference between the utilization, $UT$, and the actual conversion, $\zeta$ (the difference representing the conversion in the reforming section), first increases with $\gamma$ (starting from $\gamma=0$) but above a certain $\gamma$-value (0.5-0.75) becomes constant: $UT$ and $\zeta$ decrease simultaneously as if the conversion in the reforming section becomes independent of $\gamma$. The constant difference of $\zeta$ and $UT$ increases with $\alpha$, namely from $\zeta$-$UT$=0.17 at $\alpha$=0.42 to 0.20 at $\alpha$=0.52 and 0.23 at $\alpha$=0.59. Though the gap between $UT$ and $\zeta$ remains fairly constant for a given $\alpha$ and for enough steam (above $\gamma$~0.5-0.75), the gap between $UT$ and the thermodynamically attainable conversion increases with increasing amounts of steam, $\gamma$.

For each value of $\alpha$ ($\alpha$=0.42 in *Figure 4*a, $\alpha$=0.52 in *Figure 4*b and $\alpha$=0.59 in *Figure 4*c), thermodynamic equilibrium conversion (upper solid line) was reached at the end of the reforming section for zero steam addition, $\gamma$=0. With increasing amounts of steam, the actual conversion, $\zeta$, first increases, then levels off and finally starts to decrease. This while the thermodynamically attainable conversion continuously increases with $\gamma$. As a consequence, the gap between the actual conversion, $\zeta$, and that what is thermodynamically attainable increases with $\gamma$. Interestingly, the gap between the actual conversion, $\zeta$, and thermodynamic equilibrium is about the same for each value of the oxygen-to-carbon ratio, $\alpha$, though both values strongly increase with $\alpha$.

That thermodynamic equilibrium is only reached at low to zero steam loads is due to the decreasing utilization in the oxidation section and the lower reaction rates in the reforming section when $\alpha$ is increased. The lower reaction rate is due to the lower temperatures throughout the reforming section, and the higher total gas flow rate, $\phi_{tot}$, resulting in a lower methane partial pressure. Both effects are due to the presence of the additional steam. Thus, the thermodynamic advantage of adding more steam (if available at a low cost at $T_0$) can only be attained with more catalyst (or at a lower space velocity).

As discussed, the gap between the actual, observed, conversion, and the potential, thermodynamically attainable, conversion does not depend much on the oxygen-to-carbon ratio, $\alpha$. This shows that several opposing mechanisms occur simultaneously: though reactor temperatures are much higher at the high air-load (high value for $\alpha$), the methane partial pressure is decreased due to the additional nitrogen, while the actual conversion required in the reforming section (to keep the gap constant) increases with $\alpha$ (from ~0.17 at $\alpha$=0.42 to ~0.23 at $\alpha$=59).

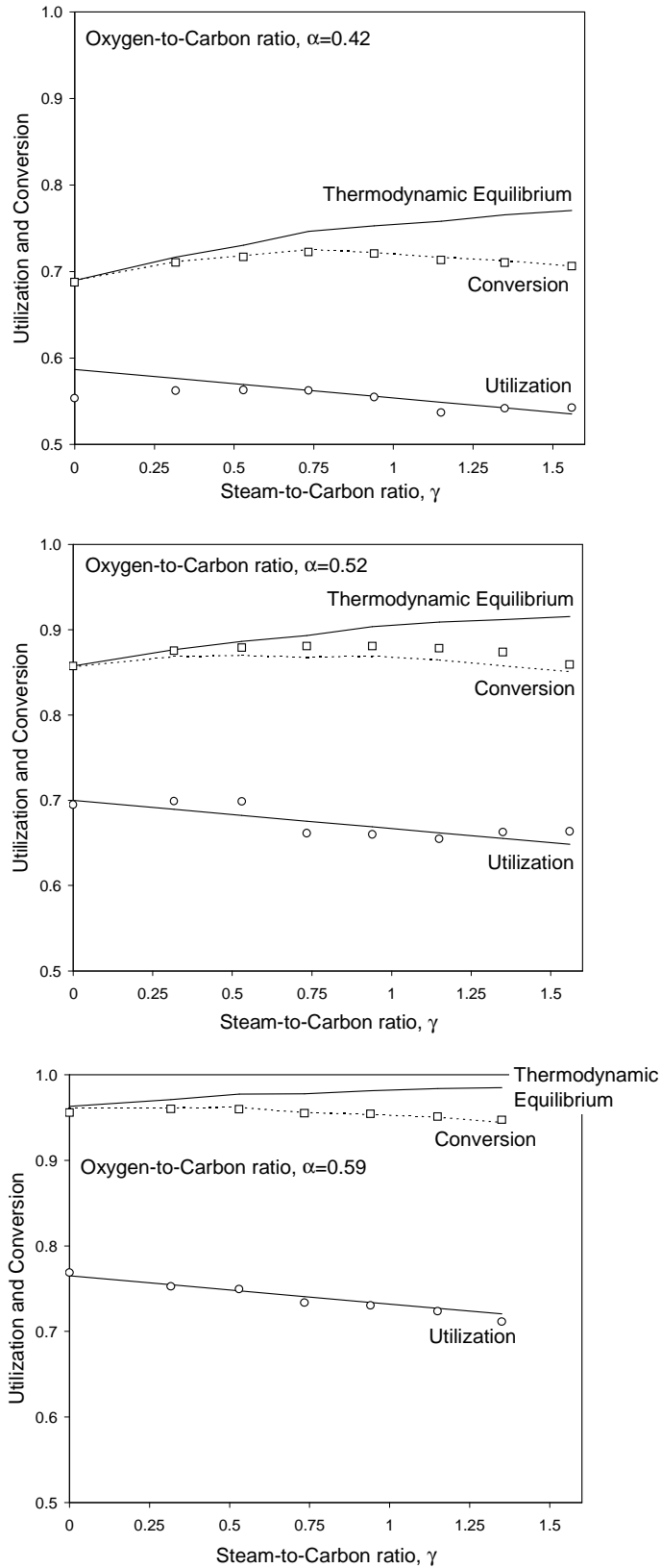

*Figure 4.* Methane utilization and conversion. The fuel utilization ratio, *UT*, equals the methane conversion at the end of the oxidation section and is derived from top-temperature data (circles). The lower solid line is the empirical fit, Eq. [6]. The two-section reactor model is used for the methane conversion, $\zeta$, at the end of the reforming section (dashed line) and is compared with conversion data (squares). Upper solid line is thermodynamic equilibrium. $T_0 \sim 400$ °C, $P=170$ kPa, $E_A=160$ kJ/mol, $\lambda=17.26$ mol/(Pa·s), $\phi_{CH4,0}=5.3$ mmol/s, $\kappa=0$ W/K, varying heat loss $Q_{ox}$).

Further Simulations

For other parameter settings, simulation results are presented in *Figure 5 - Figure 7*. *Figure 5* focuses on the oxidation section and shows the top-temperature and methane utilization as function of the oxygen-to-carbon ratio, $\alpha$. The top-temperature is plotted on the left-hand y-axis and increases with more oxygen and less steam. The utilization is plotted on the right-hand y-axis (dashed lines) and as in *Figure 4* increases with more oxygen and less steam. Again, as in *Figure 4* the maximum, thermodynamically attainable, conversion (solid lines) increases with more steam and oxygen. For the pressure of 150 kPa, only for sufficient amounts of oxygen (in this case, for $\alpha$>0.6) is the thermodynamic methane conversion close to 100%. For a lower $\alpha$, the thermodynamic, maximum, fuel conversion is significantly reduced and most significantly without steam ($\gamma$=0).

*Figure 5* indicates once more that with more steam the utilization (conversion in oxidation section) decreases while the attainable (maximum) conversion increases: more fuel must be converted in the reforming section.

Simulation results of the development of temperature and methane conversion through the reforming section are presented in *Figure 6*. Methane conversion is directly coupled to syngas yield using Eq. [16]. The top-temperature is higher without steam which is one reason why conversion is already complete at $L^*$~0.3. For $\gamma$=1.6 methane is converted along the entire reactor coordinate, with thermodynamic equilibrium not yet reached at $L^*$=1.

Comparing $\gamma$=0 and $\gamma$=1.6 shows that for $\gamma$=1.6 more catalyst is required to arrive at thermodynamic equilibrium, but that the methane conversion and syngas yield associated with equilibrium are higher (+1.5% and +1%, respectively) than at zero steam load.

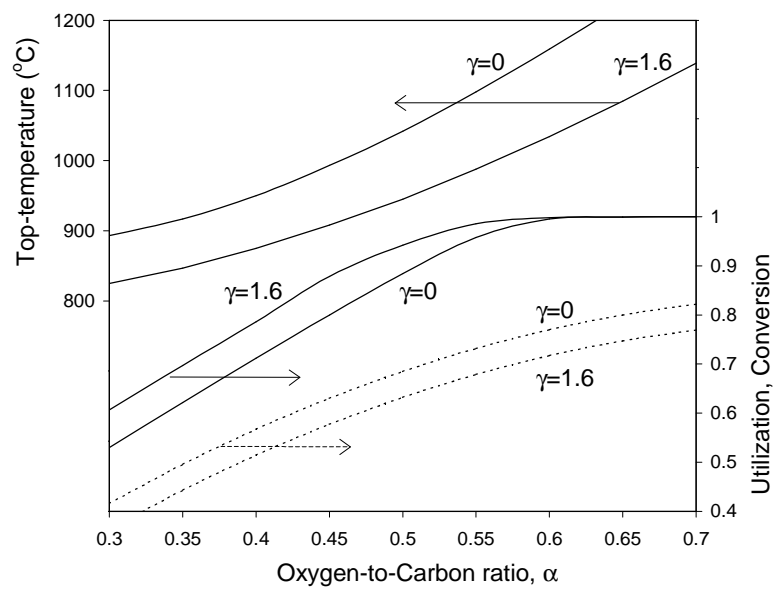

*Figure 5*. Oxidation Section in ATR/CPO of methane. Variation of oxygen-to-carbon ratio, $\alpha$, and steam-to-carbon ratio, $\gamma$. Top-temperature on left y-axis. Thermodynamic conversion (solid lines on right y-axis) as well as the fuel utilization (dashed lines). $T_0$=400 $^\circ$C, $Q_{ox}$=0 W, $P$=150 kPa.

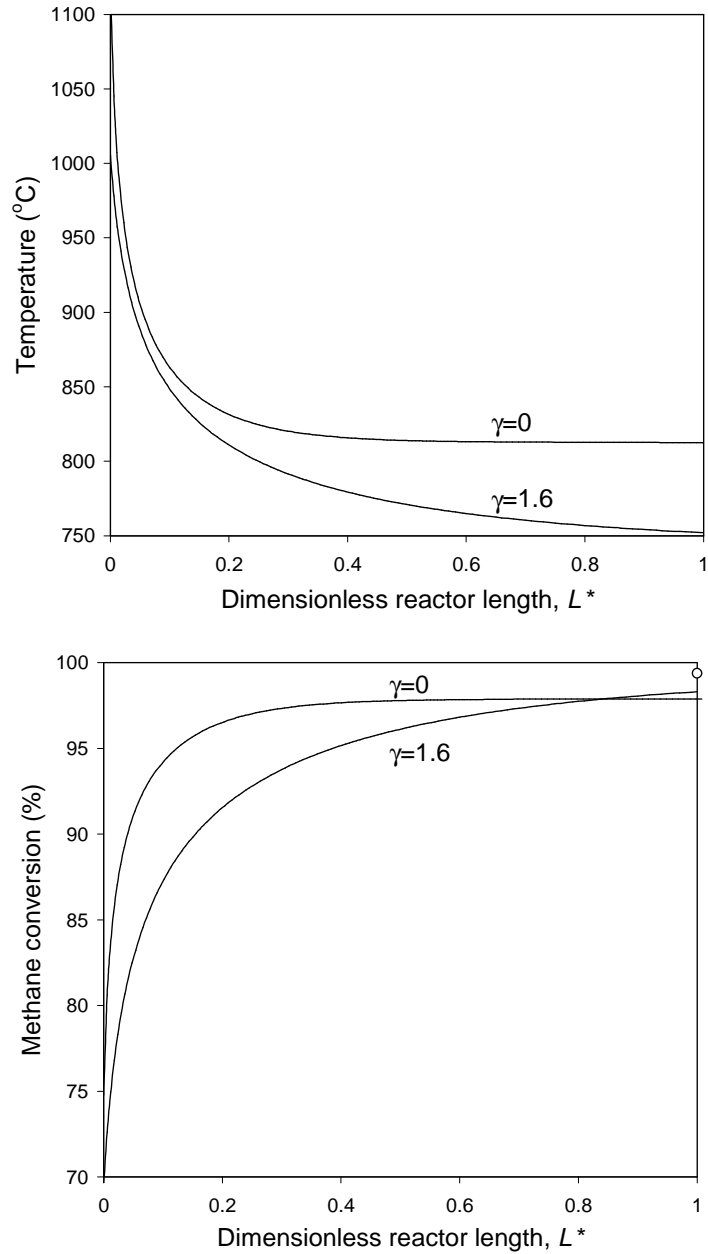

*Figure 6.* Temperature profile (a) and methane conversion (b) as function of reactor coordinate, $L^*$, for two steam-to-carbon ratio's, $\gamma$. The circle shows the equilibrium value for $\gamma$=1.6 ($\alpha$=0.57, $P$=150 kPa, $\phi_{CH4,0}$= 10 mmol/s, $T_0$=400 $^o$C, $\lambda$=0.10 mol/(Pa·s), $E_A$=100 kJ/mol, $Q_{ox}$=0 W, $\kappa$=0 W/K).

In case of heat loss through the reactor wall, $\kappa$>0, the situation becomes more complicated and it is possible that for a low $\gamma$, resulting in fast reactions and low catalyst volumes, the optimum $L^*$ corresponds to a higher methane conversion (thus, a higher syngas yield). *Figure 7* shows that for $\gamma$=0 the maximum methane conversion is found at $L^*$~0.14 and is 95.7%, whereas for $\gamma$=1.6 the maximum methane conversion is 91.5% at $L^*$~0.63. This behavior is opposite to what is predicted for thermodynamic equilibrium and without heat loss, in which case methane conversion is higher for $\gamma$=1.6 than for $\gamma$=0, see *Figure 6*. In the decreasing portions of the conversion curves (*Figure 7*) thermodynamic equilibrium has already been rather well-established but due to heat loss the temperature decreases and methane is formed again (methanation).

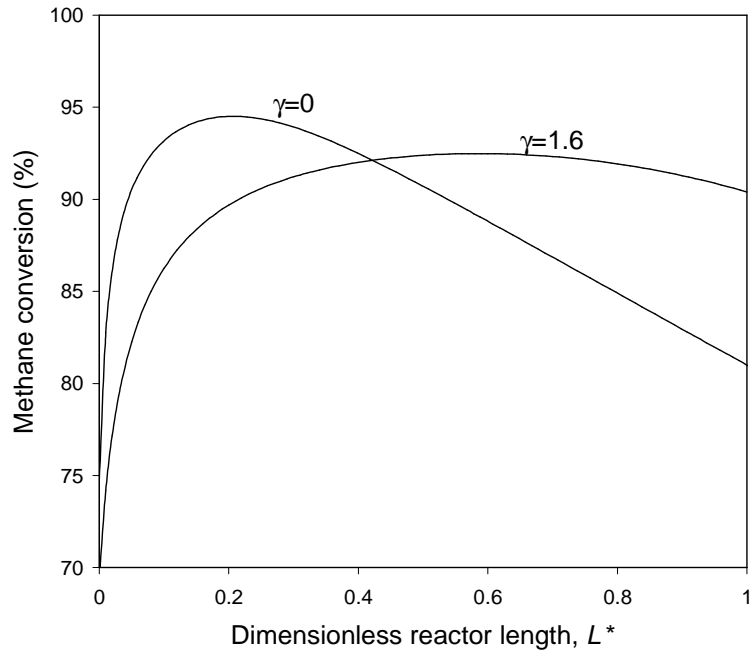

*Figure 7.* Methane conversion with heat loss in the reforming section, as function of $\gamma$ and $L^*$. ($\kappa$=1 W/K, $T^*$=298.15 K, all other data see *Figure 6*).

## Conclusions

A two-section reactor model describes several key features of fuel conversion in autothermal reforming (CPO and ATR) based on air and steam, such as the high upstream temperature as well as the decreasing temperature throughout the reforming section. The model can be fitted to data to obtain an optimum value for the kinetic constant, activation energy and reaction order. For a given amount of catalyst and with increasing steam loads, the temperature in the reactor decreases while the methane conversion (and yield of hydrogen and CO) first increases and then decreases, while the conversion that is thermodynamically feasible continuously increases. Thus, the thermodynamic advantage of adding more steam can only be attained at a lower space velocity. The model can be extended to describe multicomponent fuels such as natural gas and naphtha.

## Notation

| | | |
|---|---|---|
| $a$ | Specific surface area of catalyst | m$^2$/m$^3$ |
| $c_i$ | Concentration of component i | mol/m$^3$ |
| $E_A$ | Activation energy of reaction i | J/mol |
| $H_i$ | Enthalpy of component i | J/mol |
| $k_0$ | Pre-exponential kinetic number | mol/(m$^2$.Pa.s) |
| $K_i$ | Equilibrium constant of reaction i (either WGS or MSR) | |
| $L^*$ | Dimensionless axial reactor coordinate | |
| $p_i$ | Partial pressure of component i | Pa |
| $P$ | Absolute pressure | Pa |
| $P^*$ | Reference pressure (=1.013·10$^5$) | Pa |
| $P_R$ | Perimeter of the reactor | m |
| $Q_{ox}$ | Heat loss in oxidation section | W |
| $R$ | Gas constant (=8.3144) | J/(mol.K) |
| $R_i$ | Reaction rate of reaction i | mol/(m$^3$.s) |
| $S$ | Cross-sectional surface area of reactor | m$^2$ |
| $S_i$ | Entropy of component i | J/(mol.K) |
| $t$ | Time | s |
| $T$ | Temperature | K |
| $T_0$ | Inlet temperature | K |
| $T^*$ | Temperature outside of reactor | K |
| $UT$ | Fuel utilization ratio | |
| $U$ | Effective heat transfer coefficient | W/(m$^2$.K) |
| $V$ | Reactor volume | m$^3$ |
| $x_i$ | mole fraction of component i | mol/mol |
| $z$ | Reactor coordinate in axial direction | m |

*Greek symbols*

| | | |
|---|---|---|
| $\alpha$ | Oxygen-to-carbon ratio of inlet flow (=$\phi_{O2,0}/\phi_{CH4,0}$) | |
| $\gamma$ | Steam-to-carbon ratio of inlet flow (=$\phi_{H2O,0}/\phi_{CH4,0}$) | |
| $\zeta$ | Methane conversion (at the end of reforming section) | |
| $\kappa$ | =$U \cdot \pi \cdot V/S$ | W/K |
| $\lambda$ | =$a \cdot k_0 \cdot V$ | mol/(Pa.s) |
| $\nu_i$ | Stoichiometric constant of reactant i | |
| $\phi_i$ | Mole flows of component i | mol/s |
| $\phi_{i,0}$ | Inlet mole flows of component i | mol/s |
| $\phi_{tot}$ | Total mole flow | mol/s |
| $\phi_{i,top}$ | Mole flow of component i at the intersection of oxidation and reforming section | mol/s |
| $\chi$ | Fraction of oxygen in air (=0.2095) | |

## Literature Cited

Ahmed, S., and M. Krumpelt, "Hydrogen from hydrocarbon fuels for fuel cells," *Int. J. hydrogen energy*, **26** (2001) 291.

Akers, W.W., and D.P. Camp, "Kinetics of the methane-steam reaction," *AIChE J.*, **1**, 471 (1955).

Arena, F., F. Frusteri, and A. Parmaliana, "Kinetics of the partial oxidation of methane to formaldehyde on silica catalyst," *AIChE J.*, **46**, 2285 (2000).

Bodke, A.S., S.S. Bharadwaj, and L.D. Schmidt, "The effect of ceramic supports on partial oxidation of hydrocarbons over noble metal coated monoliths," *J. Catalysis*, **179**, 138 (1998).

Bodrov, N.M., L.O. Apel'baum, and M.I. Temkin, "Kinetics of the reaction of methane with steam on the surface of nickel," *Kinet. Catal.*, **5**, 614 (1964).

Chang, Y.-F., and H. Heinemann, "Partial oxidation of methane to syngas over Co/MgO catalysts. Is it low temperature?," *Catalysis Letters*, **21**, 215 (1993).

Daubert, T.E., and R.P. Danner, *Data Compilation Tables of properties of pure compounds*, AIChE, New York (1985).

De Groote, A., and G.F. Froment, "Reactor modeling and simulations in synthesis gas production," *Reviews in Chem. Eng.*, **11**, 145 (1995).

De Groote, A., and G.F. Froment, "Simulation of the catalytic partial oxidation of methane to synthesis gas," *Applied Catalysis A: General*, **138**, 245 (1996).

De Smet, C.R.H., M.H.J.M. de Croon, R.J. Berger, G.B. Marin, and J.C. Schouten, "Kinetics for the partial oxidation of methane on a Pt gauze at low conversions," *AIChE J.*, **46**, 1837 (2000).

De Smet, C.R.H., M.H.J.M. de Croon, R.J. Berger, G.B. Marin, and J.C. Schouten, "Design of adiabatic fixed bed reactors for the partial oxidation of methane to synthesis gas. Application to production of methanol and hydrogen-for-fuel-cells," *Chem. Eng. Sci.*, **56**, 4849 (2001).

Deutschmann, O., and L.D. Schmidt, "Modeling the partial oxidation of methane in a short-contact-time reactor," *AIChE J.*, **44**, 2465 (1998).

Dissanayake, D., M.P. Rosynek, K.C.C. Kharas, and J.H. Lunsford, "Partial oxidation of methane to carbon monoxide and hydrogen over a Ni/Al$_2$O$_3$ catalyst," *J. Catalysis*, **132**, 117 (1991).

Docter, A., and A. Lamm, "Gasoline fuel cell systems," *J. Power Sources*, **84**, 194 (1999).

B. Elvers *et al.* (ed.). *Ullmann's Encyclopedia of Industrial Chemistry*, Volume A12, p. 207, 5th Ed., VCH, Weinheim (1989).

Heitnes Hofstad, K., T. Sperle, O.A. Rokstad, and A. Holmen, "Partial oxidation of methane to synthesis gas over a Pt/10% Rh gauze," *Catalysis Letters*, **45**, 97 (1997).

Hickman, D.A., and L.D. Schmidt, "Production of syngas by direct catalytic oxidation of methane," *Science*, **259**, 343 (1993a).

Hickman, D.A., and L.D. Schmidt, "Steps in CH$_4$ oxidation on Pt and Rh surfaces: high temperature reactor simulations, " *AIChE J.*, **39**, 1164 (1993b).

Hou, K., and R. Hughes, "The kinetics of methane steam reforming over a Ni/$\alpha$-Al$_2$O catalyst," *Chem. Eng. J.*, **82**, 311 (2001).

Kramer, G.J., W. Wieldraaijer, P.M. Biesheuvel, and H.P.C.E. Kuipers, "The determining factor for catalyst selectivity in Shell's catalytic partial oxidation process," *ACS Fuel Chemistry Division Preprints*, **46**, 659 (2001).

Moon, D.J., K. Sreekumar, S.D. Lee, B.G. Lee, and H.S. Kim, "Studies on gasoline fuel processor system for fuel-cell powered vehicles application," *Applied catalysis A: General*, **215**, 1 (2001).

Papp, H., P. Schuler, and Q. Zhuang, "CO$_2$ reforming and partial oxidation of methane," *Topics in Catalysis*, **3**, 299 (1996).

Prettre, M., Ch. Eichner, and M. Perrin, "The catalytic oxidation of methane to carbon monoxide and hydrogen," *Trans. Faraday Soc.*, **43**, 335 (1946).

Ridler, D.E., and M.V. Twigg, "Steam Reforming." Ch. 5 in *Catalyst Handbook*. M.V. Twigg (Ed.), Manson Publishing, London (1989).

Rostrup-Nielsen, J.R., "Production of synthesis gas," *Catalysis Today*, **18**, 305 (1993).

Service, R.F., "Bringing fuel cells down to earth," *Science*, **285**, 682 (1999).

Soliman, M.A., A.M. Adris, A.S. Al-Ubaid, and S.S.E.H. El-Nashaie, "Intrinsic kinetics of nickel/calcium aluminate catalyst for methane steam reforming," *J. Chem. Tech. Biotechnol.*, **55**, 131 (1992).

Steghuis, A.G., J.G. van Ommen, and J.A. Lercher, "On the reaction mechanism for methane partial oxidation over yttria/zirconia," *Catalysis Today*, **46**, 91 (1998).

Wang, H.Y., and E. Ruckenstein, "Catalytic partial oxidation of methane to synthesis gas over γ-$Al_2O_3$-supported rhodium catalysts," *Catalysis Letters*, **59**, 121 (1999).

Zhu, J., D. Zhang, and K.D. King, "Reforming of $CH_4$ by partial oxidation: thermodynamic and kinetic analyses," *Fuel*, **80**, 899 (2001).

## Appendix A

Enthalpy, $H_i$, and entropy, $S_i$, are expressed as (Daubert and Danner, 1985):

$$H_i = H_{0,i} + a_i(T - T_0) + b_i c_i \left( \coth \frac{c_i}{T} - \coth \frac{c_i}{T_0} \right) - d_i e_i \left( \tanh \frac{e_i}{T} - \tanh \frac{e_i}{T_0} \right),$$

$$S_i = S_{0,i} + a_i \ln \frac{T}{T_0} + b_i \left( \frac{c_i}{T} \coth \frac{c_i}{T} - \frac{c_i}{T_0} \coth \frac{c_i}{T_0} - \ln \left( \frac{\sinh\left(\frac{c_i}{T}\right)}{\sinh\left(\frac{c_i}{T_0}\right)} \right) \right) - d_i \left( \frac{e_i}{T} \tanh \frac{e_i}{T} - \frac{e_i}{T_0} \tanh \frac{e_i}{T_0} - \ln \left( \frac{\cosh\left(\frac{e_i}{T}\right)}{\cosh\left(\frac{e_i}{T_0}\right)} \right) \right)$$

with $T_0$=298.15 K and coth=1/tanh. The values for $a_i$-$e_i$ are summarized in *Table A*.

*Table A*. Thermodynamic data.

|  | $H_{0,i}$ (kJ/mol) | $S_{0,i}$ (J/mol/K) | $a_i$ (J/mol/K) | $b_i$ (J/mol/K) | $c_i$ (K) | $d_i$ (J/mol/K) | $e_i$ (K) |
|---|---|---|---|---|---|---|---|
| $CH_4$ | -74.5 | 186.27 | 33.298 | 79.933 | 2086.9 | 41.602 | 991.96 |
| $O_2$ | 0 | 205.04 | 29.103 | 10.04 | 2526.5 | 9.356 | 1153.8 |
| CO | -111 | 197.54 | 29.108 | 8.773 | 3085.1 | 8.4553 | 1538.2 |
| $CO_2$ | -394 | 213.69 | 29.37 | 34.54 | -1428 | 26.4 | 588 |
| $H_2$ | 0 | 130.57 | 27.617 | 9.56 | 2466 | 3.76 | 567.6 |
| $H_2O$ | -242 | 188.72 | 33.363 | 26.79 | 2610.5 | 8.896 | 1169 |
| $N_2$ | 0 | 191.609 | 29.105 | 8.6149 | 1701.6 | 0.10347 | 909.79 |